\documentclass[aip,apl,numerical,a4,reprint,twoside,floatfix,superscriptaddress]{revtex4-1} 
\usepackage{graphicx}
\usepackage{bm}
\usepackage{color}
\usepackage{mathrsfs}

\begin{document}
\title[]{Angular momentum compensation manipulation to room temperature of the ferrimagnet Ho$_{3-x}$Dy$_{x}$Fe$_5$O$_{12}$ detected by the Barnett effect}
\author{Masaki Imai}
\email{imai.masaki@jaea.go.jp}
\affiliation{Advanced Science Research Center, Japan Atomic Energy Agency, Tokai 319-1195, Japan}
\author{Hiroyuki Chudo}
\affiliation{Advanced Science Research Center, Japan Atomic Energy Agency, Tokai 319-1195, Japan}
\author{Masao Ono}
\affiliation{Advanced Science Research Center, Japan Atomic Energy Agency, Tokai 319-1195, Japan}
\author{Kazuya Harii}
\affiliation{Advanced Science Research Center, Japan Atomic Energy Agency, Tokai 319-1195, Japan}
\author{Mamoru Matsuo}
\affiliation{Advanced Science Research Center, Japan Atomic Energy Agency, Tokai 319-1195, Japan}
\affiliation{Kavli Institute for Theoretical Sciences, University of Chinese Academy of Sciences,19 Yuquan Road, Beijing 100049, P.R.China}
\affiliation{Riken Center for Emergent Matter Science (CEMS), Wako 351-0198, Japan}
\author{Yuichi Ohnuma}
\affiliation{Kavli Institute for Theoretical Sciences, University of Chinese Academy of Sciences,19 Yuquan Road, Beijing 100049, P.R.China}
\affiliation{Riken Center for Emergent Matter Science (CEMS), Wako 351-0198, Japan}
\author{Sadamichi Maekawa}
\affiliation{Advanced Science Research Center, Japan Atomic Energy Agency, Tokai 319-1195, Japan}
\affiliation{Riken Center for Emergent Matter Science (CEMS), Wako 351-0198, Japan}
\affiliation{Kavli Institute for Theoretical Sciences, University of Chinese Academy of Sciences,19 Yuquan Road, Beijing 100049, P.R.China}
\author{Eiji Saitoh}
\affiliation{Advanced Science Research Center, Japan Atomic Energy Agency, Tokai 319-1195, Japan}%
\affiliation{Advanced Institute for Materials Research, Tohoku University, Sendai 980-8577, Japan}%
\affiliation{Institute for Materials Research, Tohoku University, Sendai 980-8577, Japan}%
\affiliation{Department of Applied Physics, The University of Tokyo, Hongo, Bunkyo-ku, Tokyo, 113-8656, Japan}

\date{\today}

\begin{abstract}
We demonstrate that the angular momentum compensation temperature $T_{\rm A}$, at which the net angular momentum in the sample disappears, can be controlled in Ho$_3$Fe$_5$O$_{12}$ by partially substituting Dy for Ho.
The $T_{\rm A}$ can be detected using the Barnett effect, by which mechanical rotation magnetizes an object due to spin-rotation coupling.
We found that $T_{\rm A}$ increases with the Dy content and clarified that the $T_{\rm A}$ of Ho$_{1.5}$Dy$_{1.5}$Fe$_5$O$_{12}$ coincides with room temperature.
The Barnett effect enables us to explore materials applicable to magnetic devices utilizing the angular momentum compensation only by rotating the powder sample at room temperature.
\end{abstract}

\maketitle

Angular momentum compensation is a key characteristic in the field of spintronics, where attention is focused on the high-speed magnetic response at the angular momentum compensation temperature.\cite{Binder2006,Stanciu2006,Stanciu2007,Kim2017}
N-type ferrimagnets have a magnetic compensation temperature $T_{\rm M}$, at which magnetization disappears even in the ferrimagnetically ordered state.
Furthermore, when $g$-factors of the magnetic moment belonging to different sublattices are different, the ferrimagnetic materials have an additional compensation, namely the angular momentum compensation temperature $T_{\rm A}$, at which the net angular momentum $\langle J_{\rm net}\rangle $ in the material disappears even in the magnetically ordered state.
While $T_{\rm M}$ is easily determined by magnetization measurements, $T_{\rm A}$ cannot be determined by conventional magnetization measurements using a magnetic field.
It was recently reported that the domain wall mobility was enhanced at $T_{\rm A}$ in GdFeCo.\cite{Kim2017}
However, this method requires microfabrication of materials and only applies to metals.
Nevertheless, using the Barnett effect, a phenomenon by which a rotating object is magnetized by spin-rotation coupling, the angular momentum compensation can be determined.\cite{Imai2018}

The Barnett effect was discovered in 1915.\cite{Barnett1915,Barnett1935}
The angular momentum of electrons in matter interacts with rotational motion through spin-rotation coupling, described as 
\begin{equation}
{\mathscr H} = - {\bm J} \cdot { \bm \Omega},
\end{equation}
where $\bm J$ is the angular momentum and $\bm \Omega$ is the angular velocity.\cite{deOliveira1962}
The angular momentum is then aligned with the direction of the rotation axis, which causing magnetization.
With ferrimagnets, ${\bm J} $ is represented as the sum of the angular momenta belonging to different sublattices $\langle J_{\rm net}\rangle$.
The Barnett effect does not apply when $\langle J_{\rm net}\rangle=0$, which is the definition of the angular momentum compensation.

In this letter, we demonstrate $T_{\rm A}$ control for the insulating rare earth iron garnet Ho$_{3-x}$Dy$_x$Fe$_5$O$_{12}$.
$T_{\rm A}$ is observed using the Barnett effect.\cite{Imai2018}
To realize a high-speed magnetic device using the angular momentum compensation, $T_{\rm A}$ should coincide with the temperature at which the device is operated. 
We found that $T_{\rm A}$ coincides with room temperature (20$^\circ\mathrm{C}$) at $x=1.5$.

Powder samples of Ho$_{3-x}$Dy$_x$Fe$_5$O$_{12}$ were prepared from Fe$_{2}$O$_{3}$(4N), Ho$_{2}$O$_{3}$(3N), and Dy$_{2}$O$_{3}$(3N) via a solid-state reaction.
Fe$_{2}$O$_{3}$, Ho$_{2}$O$_{3}$, and Dy$_{2}$O$_{3}$ powders were mixed in a molar ratio of $5:3-x:x$ in an agate mortar.
The mixed powder was pelletized and heated to $1200\ {}^\circ\mathrm{C}-1400\ {}^\circ\mathrm{C}$ in the ambient atmosphere.
To characterize the samples, the DC-magnetization measurements were performed by  extraction magnetometry with a commercial magnetometer (PPMS, Quantum Design).

\begin{figure}
\includegraphics[width=1\linewidth]{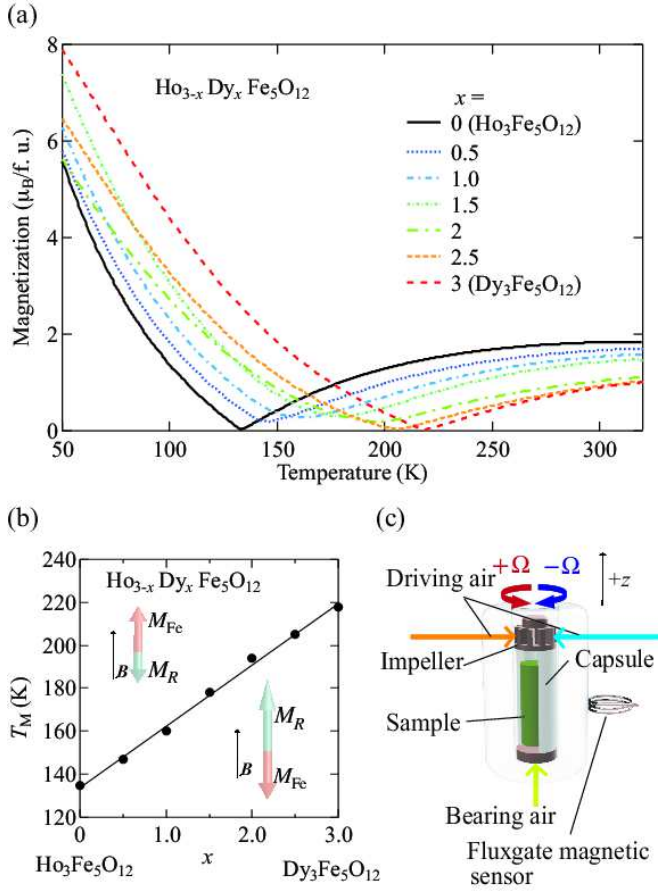}
\caption{(a) Temperature dependence of Ho$_{3-x}$Dy$_x$Fe$_5$O$_{12}$ magnetization in a magnetic field of 1,000 Oe.
(b) The variation in magnetization compensation temperature $T_{\rm M}$ with the Dy content $x$ in Ho$_{3-x}$Dy$_x$Fe$_5$O$_{12}$.
(c) Schematic of the apparatus for Barnett effect measurement using an air-driven rotor system.
\label{fig1}}
\end{figure}

Figure~\ref{fig1}(a) shows the temperature dependence of the magnetization of Ho$_{3-x}$Dy$_x$Fe$_5$O$_{12}$.
The magnetization of both the end materials, Ho$_3$Fe$_5$O$_{12}$ (HoIG) and Dy$_3$Fe$_5$O$_{12}$ (DyIG), vanishes at $T_{\rm M}$ = 135 and 218 K, respectively.
For the substituted materials, the magnetization does not disappear completely due to the $T_{\rm M}$ distribution; thus $T_{\rm M}$ was defined as the temperature at which the magnetization shows a local minimum.
Figure \ref{fig1}(b) shows the linear relation between $T_{\rm M}$ and the Dy concentration, and a schematic of the magnetization at sublattices.
In the high-temperature region above $T_{\rm M}$, since the magnetization at the Fe${^{3+}}$ sublattice, $M_{\rm Fe}$, is larger than that at the $R^{3+}$ sublattice, $M_R$, $M_{\rm Fe}$ aligns parallel to the magnetic field.
The magnitude of $M_R$ increases as the temperature decreases, and $M_R$ equals $M_{\rm Fe}$ at $T_{\rm M}$.
The magnetization of these sublattices flips across $T_{\rm M}$, and then $M_R$ aligns parallel to the magnetic field below $T_{\rm M}$ (see Section I of Supplementary Material for more details).

\begin{figure}
\includegraphics[width=1\linewidth]{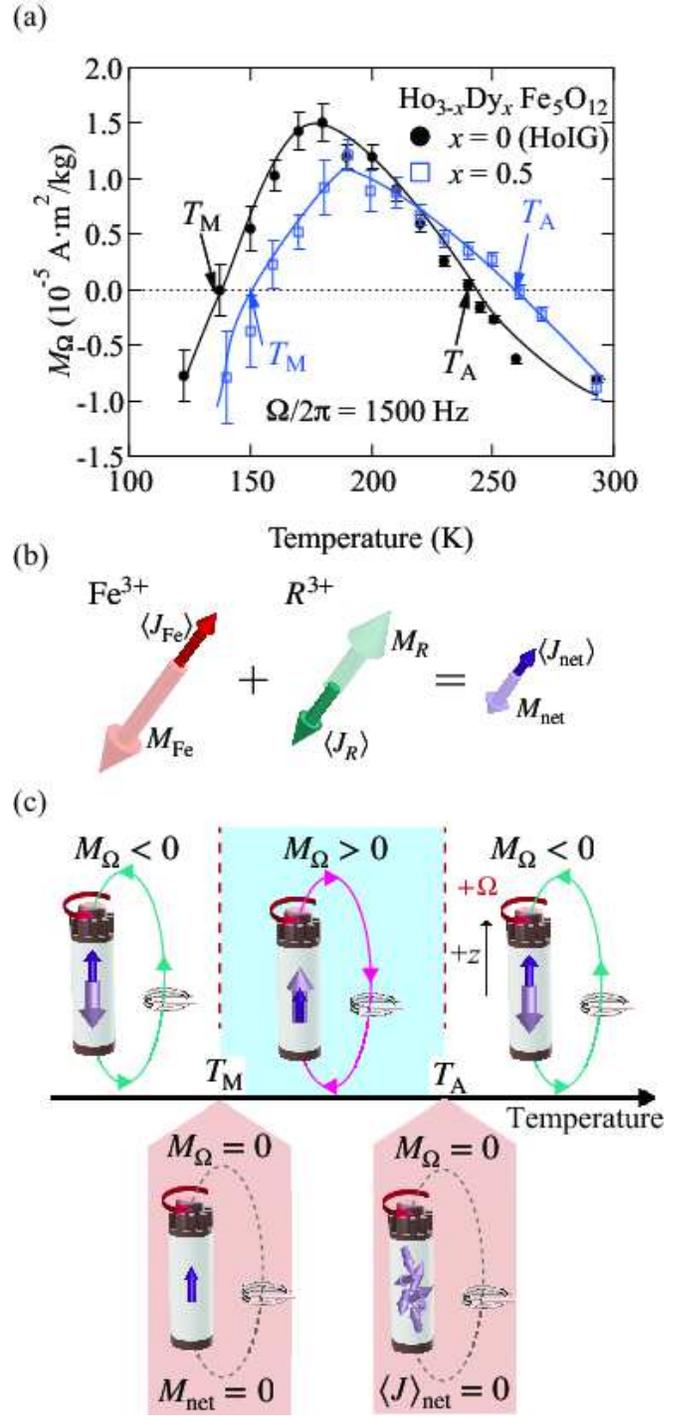}
\caption{(a) Temperature dependence of $M_\Omega$ of Ho$_{3-x}$Dy$_x$Fe$_5$O$_{12}$ ($x=0$ and $0.5$).
The solid lines are the guide for eyes.
(b) Schematic illustration of angular momentum and magnetic moment. Relationship between angular momentum (dark red and blue arrows) and magnetic moment (light red and blue arrows) in Ho$_{3-x}$Dy$_x$Fe$_5$O$_{12}$.
(c) Relationship between angular momentum and the magnetization $M_\Omega$ induced by rotation at various temperature.
The dark and light purple arrows shows net angular momentum and magnetization of Ho$_{3-x}$Dy$_x$Fe$_5$O$_{12}$, respectively.
\label{fig2}}
\end{figure}

\begin{figure*}
\includegraphics[width=1\linewidth]{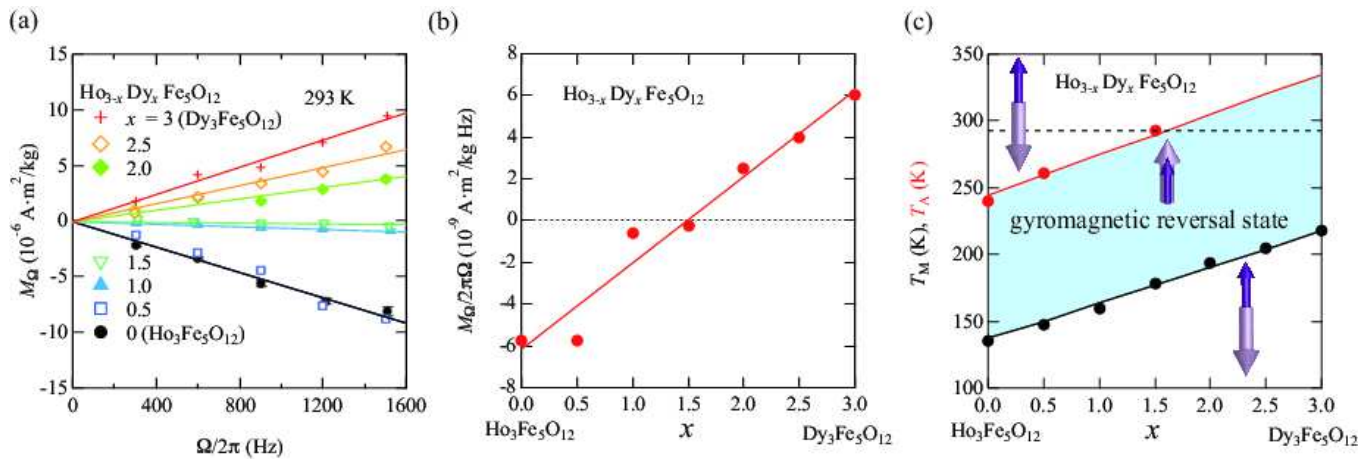}
\caption{(a) Rotational frequency dependence of $M_\Omega$ for $x=0, 0.5, 1.0, 1.5, 2.0, 2.5$ and 3.0 at 293~K.
(b) Dy content $x$ dependence of the rotation-susceptibility $M_\Omega/2\pi\Omega$ at 293~K.
(c) Phase diagram of Ho$_{3-x}$Dy$_x$Fe$_5$O$_{12}$.
Black and red circles indicate $T_{\rm M}$ and $T_{\rm A}$, respectively. 
Black line is the linear fit to the $T_{\rm M}$ data .
Red line is the calculated $T_{\rm A}$(see Section III of Supplementary Material).
The blue area represents the gyromagnetic reversal region, where the directions of the net angular momentum and net magnetic moment are aligned in parallel.
\label{fig3}}
\end{figure*}

\begin{figure}
\includegraphics[width=1\linewidth]{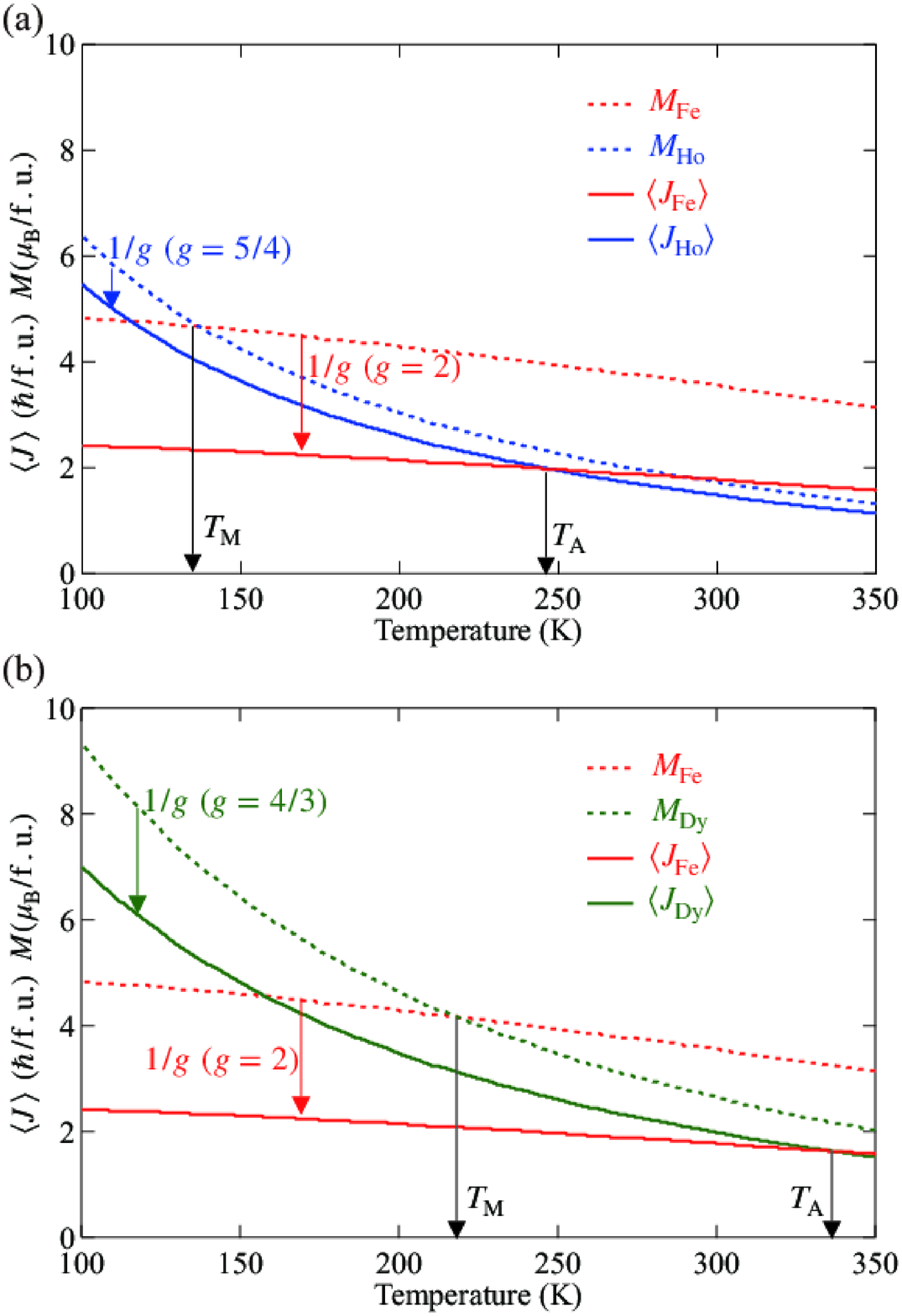}
\caption{Temperature dependence of angular momenta and magnetic moments of sublattice magnetic ions in 
Ho$_3$Fe$_5$O$_{12}$ (a) and Dy$_3$Fe$_5$O$_{12}$ (b).
Red, blue, and green dashed lines indicate the magnetic moments of Fe, Ho, and Dy sites, respectively.
Solid lines indicate the angular momenta. 
\label{fig4}}
\end{figure}

We measured the Barnett effect to observe angular momentum compensation.
Our apparatus for rotating samples uses an air-driven rotor system as shown in Fig.~\ref{fig1}(c).\cite{Ono2015,Ogata2017A,Ogata2017B,Imai2018}
The rotor system was placed in magnetic shields to exclude the geomagnetic field.
The sample was rotated using compressed air with angular velocity $\Omega$ and magnetized to $M_\Omega$ by the rotation.
We measured the stray field from $M_\Omega$ using a fluxgate magnetic sensor.
To remove the residual background magnetic field, we measured the difference in the stray field $\Delta B =[B(+\Omega)-B(-\Omega)]/2$, where $B(+\Omega)$ and $B(-\Omega)$ represent the stray fields obtained at $+\Omega$ and $-\Omega$, respectively (Fig.~\ref{fig1}(c)).
The measurements were repeated over 20 times to remove fluctuation of the background magnetic field and increase data accuracy.
$M_\Omega$ was calculated from $\Delta B$ using the dipole model described in Ref.~\onlinecite{Ogata2017A}.
For experimental details of Barnett effect measurement at room temperature and low temperatures, see Refs.~\onlinecite{Ogata2017A} and \onlinecite{Imai2018}, respectively.

Figure \ref{fig2}(a) shows the temperature dependence of $M_\Omega$ of the Dy content for $x=0$ and 0.5 at a rotational frequency of $\Omega/2\pi$=1.5~kHz.
The $M_\Omega$ of HoIG crosses zero at two temperatures, $T_{\rm M} =135$~K and $T_{\rm A}=240$~K.\cite{Imai2018}
On Dy substitution with $x=0.5$, $T_{\rm M}$ and $T_{\rm A}$ increase to 150 and 260 K, respectively. 
This indicates that Dy substitution in HoIG can control $T_{\rm M}$ and $T_{\rm A}$. 

In Ho$_{3-x}$Dy$_x$Fe$_5$O$_{12}$, the net angular momentum $\langle J_{\rm net} \rangle$ and magnetization $M_{\rm net}$ are described as $\langle J_{\rm net} \rangle =\langle J_{\rm Fe} \rangle-\langle J_R \rangle$ and $M_{\rm net}=M_{\rm Fe}-M_R$, where $\langle J_{\rm Fe} \rangle$ ($\langle J_R \rangle$) and $M_{\rm Fe}$ ($M_R$) represent the expected angular momentum and magnetization at ${\rm Fe}^{3+}$ ($R^{3+}$) sublattices, respectively (Fig.~\ref{fig2}(b)).
An effective $g$ factor is defined as $M_{\rm net}= g \langle J_{\rm net} \rangle$.

Above $T_{\rm A}$, since $\langle J_{\rm net} \rangle$ and $M_{\rm net}$ are coupled antiparallel due to the negative sign of the effective $g$ factor, $M_\Omega$ becomes negative as shown in Fig.~\ref{fig2}(c), because Eq.~(1) indicates that $+\Omega$ rotation aligns $\langle J_{\rm net} \rangle$ with the $+z$ direction.
At $T_{\rm A}$, since the Barnett effect does not apply due to the net zero angular momentum, $M_\Omega$ becomes zero.
In the range of temperatures between $T_{\rm A}$ and $T_{\rm M}$, $M_\Omega$ becomes positive.
This indicates that $\langle J_{\rm net} \rangle$ and $M_{\rm net}$ are coupled parallel; the sign of the $g$ factor becomes positive: this is a gyromagnetic reversal state.\cite{Imai2018}
At $T_{\rm M}$, though $\langle J_{\rm net} \rangle$ aligns with the rotation axis, $M_\Omega$ is zero due to disappearance of $M_{\rm net}$ at $T_{\rm M}$.
Below $T_{\rm M}$, $M_\Omega$ becomes negative again due to the antiparallel coupling between $\langle J_{\rm net} \rangle$ and $M_{\rm net}$.
The increase in $T_{\rm A}$ with increasing $x$ indicates that $T_{\rm A}$ coincide with room temperature (293 K) at appropriate Dy substitution.

To investigate the Ho$_{3-x}$Dy$_x$Fe$_5$O$_{12}$ composition with $T_{\rm A}=293$~K, the rotational frequency dependence of $M_\Omega$ was measured for various Ho$_{3-x}$Dy$_x$Fe$_5$O$_{12}$ compositions at 293~K as shown in Fig.~\ref{fig3}(a).
For all compositions, $M_\Omega$ is proportional to $\Omega$, indicating that the Barnett effect was correctly measured.
In HoIG, $M_\Omega$ is negative and decreases as $\Omega$ increases because $\langle J_{\rm net} \rangle$ and $M_{\rm net}$ are coupled antiparallel above $T_{\rm A} =240$~K (see Fig.\ref{fig2}(a)).
As the Dy content $x$ increases, the slope of $M_\Omega$ vs. $\Omega$ gradually increases, becoming almost 0 at $x=1.5$, and becoming positive at $x \geq 2.0$.
Figure \ref{fig3}(b) shows the initial susceptibility induced by rotation: $M_\Omega/2\pi\Omega$ at 293~K.
$M_\Omega/2\pi\Omega$ increases with increasing Dy content $x$. 
From the linear fit to the data, $T_{\rm A}$ is equal to 293~K at $x =1.49 \pm 0.31$.

Figure \ref{fig3}(c) shows a phase diagram of Ho$_{3-x}$Dy$_x$Fe$_5$O$_{12}$.
Both $T_{\rm M}$ and $T_{\rm A}$ depend linearly on $x$.
The black line is the linear fit to the $T_{\rm M}$ data.
The red line is obtained by calculating the temperature dependence of $\langle J_R\rangle$ with various Dy contents (see Sections II and III of Supplementary Material).
The calculations are consistent with our experimental results.

Here, we consider the Dy substitution effect on $T_{\rm M}$.
In the general rare earth iron garnet ($R$IG) system, $T_{\rm M}$ predominantly depends on $M_R$, which contains two factors: the size of the $R^{3+}$ magnetic moment $\mu=-gJ$, and $R^{3+}$—Fe$^{3+}$ interaction (see Section I of Supplementary Material).
Since the N\'eel temperature is almost irrelevant to the rare earth $R^{3+}$, the ferrimagnetic order in $R$IG is predominantly contributed by the Fe$^{3+}$—Fe$^{3+}$ interaction.
Thus, the temperature dependence of magnetization at the Fe$^{3+}$ sublattice can be regarded as almost the same in the $R$IG series.
The $R^{3+}$—Fe$^{3+}$ interaction contributes to the temperature dependence of the magnetization at $R^{3+}$ sublattice.
The $R^{3+}$—$R^{3+}$ interaction is much weaker than any other interactions in $R$IG and is negligible except in the very low temperature region.
With HoIG and DyIG, since Ho$^{3+}$ and Dy$^{3+}$ have the same magnetic moment of $\mu= 10\ \mu_{\rm B}$, $T_{\rm M}$ only depends on the $R^{3+}$—Fe$^{3+}$ interaction.
Figure \ref{fig4} shows the calculated temperature dependence of sublattice magnetization (see Section II of Supplementary Material).
Because of the stronger Dy$^{3+}$—Fe$^{3+}$ interaction compared to Ho$^{3+}$—Fe$^{3+}$, the magnetization at the Dy$^{3+}$ sublattice is larger than that at the Ho$^{3+}$ sublattice.
Consequently, the $T_{\rm M}$ of DyIG is higher than that of HoIG.

In contrast to $T_{\rm M}$, $T_{\rm A}$ depends on $\langle J_R \rangle$, which contains two factors: the size of the $R^{3+}$ angular momentum $J_R$, and the $R^{3+}$—Fe$^{3+}$ interaction.
The calculated temperature dependence of the angular momenta at the ${\rm Ho}^{3+}$ and ${\rm Dy}^{3+}$ sublattices is shown in Fig.~\ref{fig4}.
Although the angular momentum of Dy$^{3+}$ ($J_{\rm Dy}=15/2$) is smaller than that of Ho$^{3+}$ ($J_{\rm Ho}=8$), $\langle J_{\rm Dy} \rangle$ is larger than $\langle J_{\rm Ho} \rangle$ because the Dy$^{3+}$—Fe$^{3+}$ interaction is stronger than the Ho$^{3+}$—Fe$^{3+}$ interaction.
Therefore, the $T_{\rm A}$ of DyIG is larger than that of HoIG.

We now discuss the relative values of $T_{\rm M}$ and $T_{\rm A}$.
Essentially, the difference between $T_{\rm M}$ and $T_{\rm A}$ originates in the difference in the $g$ factors of the ions belonging to different sublattices.
Especially, when all the ions have the same $g$ factor, $T_{\rm A}$ merges with $T_{\rm M}$. 
When the difference in $g$ factors between Fe$^{3+}$ and $R^{3+}$ increases, so does the difference between $T_{\rm M}$ and $T_{\rm A}$.
In the Ho$_{3-x}$Dy$_x$Fe$_5$O$_{12}$ system, the $g$ factors of Fe$^{3+}$, Ho$^{3+}$, and Dy$^{3+}$ are 2, 5/4, and 4/3, respectively.
Since the difference in the $g$ factors in DyIG is smaller than that in HoIG, the difference between $T_{\rm M}$ and $T_{\rm A}$ is expected to be smaller in DyIG than in HoIG.
However, the experimental result shows that $T_{\rm A}(x)$ increases at almost the same rate as $T_{\rm M}(x)$ against temperature as shown in Fig.\ref{fig3}(c), 
This is due to competition between two effects of Dy substitution: $T_{\rm A}$ increases due to the stronger Dy$^{3+}$—Fe$^{3+}$ interaction compared to Ho$^{3+}$—Fe$^{3+}$, and $T_{\rm A}$ decreases relative to $T_{\rm M}$ due to the larger $g$ factor of Dy$^{3+}$ compared to that of Ho$^{3+}$.

In summary, we controlled the angular momentum compensation temperature of Ho$_{3-x}$Dy$_x$Fe$_5$O$_{12}$ by Dy substitution, and found that $T_{\rm A}$ coincides with room temperature at $x=1.5$ in this system using our apparatus for measuring the Barnett effect.
Using the Barnett effect, $T_{\rm A}$ is easily obtained without microfabrication of the sample, regardless of the metal or insulator. 
The Barnett effect enables us to explore candidate materials for high-speed magnetic devices exploiting fast-magnetization reversal at angular momentum compensation.

\section*{Supplementary material}
See supplementary material for details of the magnetic structure of the samples and their calculated magnetization and angular momentum. 

\begin{acknowledgments}
This work was supported by JST ERATO Grant Number JPMJER1402, JSPS Grant-in-Aid for Scientific Research on Innovative Areas Grant Number JP26103005,
and JSPS KAKENHI Grant Numbers JP16H04023, JP17H02927.
\end{acknowledgments}

\end{document}